\documentstyle[12pt]{article}
\newcommand{\la}{\label}

\newcommand{\non}{\nonumber}
\newcommand{\be}{\begin{equation}}
\newcommand{\ee}{\end{equation}}
\newcommand{\ba}{\begin{eqnarray}}
\newcommand{\ea}{\end{eqnarray}}
\newcommand{\bastar}{\begin{eqnarray*}}
\newcommand{\eastar}{\end{eqnarray*}}
\begin{document}

\begin{titlepage}
 
\vskip 4.0cm
 
\begin{center}
{\bf \large ABELIAN DECOMPOSITION OF \\ \vskip 0.3cm
                SO(2N) YANG-MILLS THEORY \\}
\end{center}
 
\vskip 2.0cm
 
\begin{center}
{\bf Wang-Chang Su$^{*}$ }  \\
\vskip 0.3cm
{\it Department of Physics \\
     National Chung-Cheng University, Chia-Yi, Taiwan } \\
\end{center}
 
\vskip 0.5cm

{\rm
Faddeev and Niemi have proposed a decomposition of $SU(N)$ Yang-Mills theory in terms of new variables, appropriate for describing the theory in the infrared limit. 
We extend this method to $SO(2N)$ Yang-Mills theory. 
We find that the $SO(2N)$ connection decomposes according to irreducible representations of $SO(N)$. 
The low energy limit of the decomposed theory is expected to describe soliton-like configurations with nontrivial topological numbers. 
How the method of decomposition generalizes for $SO(2N+1)$ Yang-Mills theory is also discussed.
}

\noindent\vfill
 
\begin{flushleft}
\rule{5.1 in}{.007 in} \\
$^{*}$ \hskip 0.2cm {\small  E-mail: \bf suw@phy.ccu.edu.tw } \\
\end{flushleft}

\end{titlepage}

The mechanism of color confinement in Yang-Mills theory is known to be one of most difficult problems in theoretical physics. 
A qualitative explanation to this problem is provided by the monopole condensation that causes the confinement of color through the dual Meissner effect \cite{nambu,thooft}.
It is conjectured that an Abelian projection of the Yang-Mills theory to its maximal Abelian subgroup is responsible for the dynamics of the dual Meissner effect.
That is to say, in the infrared limit the degrees of freedom of a non-Abelian theory are dominated by those of its maximal Abelian subgroup \cite{thooft}.
However, a quantitative understanding on how the monopole condenses in the low energy limit, starting from the fundamental Yang-Mills theory is still absent and awaits to be explored.

Recently Faddeev and Niemi had proposed an Abelian decomposition of the four-dimensional $SU(2)$ Yang-Mills connection $A^a_\mu$ \cite{fad0}. 
The decomposed theory, which is appropriate for describing the Yang-Mills theory in its infrared limit, involves an Abelian gauge field $C_\mu$, a complex scalar field $\phi = \rho + i \sigma$, and a three component unit vector field $n^a$. 
It is an on-shell decomposition because the variations of the decomposed theory with respect to the fields \( (C_\mu, \phi, n^a) \) reproduce the equations of motion of the original $SU(2)$ Yang-Mills theory. 
It is shown, on one hand, that if the fields \( (C_\mu, \phi) \) are properly integrated out, the resultant theory supports string-like knotted solution, which describes at large distance the dynamics of extended, massive knotlike solitons \cite{fad1}.
These solitonic configurations can be regarded as the natural candidates for describing glueballs. 
On the other hand, if the vector field $n^a$ is averaged over first, the multiplet \( (C_\mu, \phi) \) transforms as the fields in the Abelian Higgs model.

The method of Abelian decomposition based on $SU(2)$ Yang-Mills theory is readily generalized to the general case of four-dimensional $SU(N)$ Yang-Mills theory \cite{periwal,fad2}.
It is found that the $SU(N)$ connection decomposes according to irreducible representations of $SO(N-1)$ and that the low energy limit of the decomposed theory may describe stable, soliton-like configurations with nontrivial topological numbers \cite{fad2}.
The first principle derivation on the effective action that describes the Yang-Mills theories in the infrared limit can be found in \cite{cho1,lang,li}.

In this letter, we extend the Abelian decomposition of four-dimensional $SU(N)$ Yang-Mills connection to the case of $SO(2N)$ gauge theory. 
We shall construct the $N$ mutually orthogonal Lie-algebra valued vector fields $m_i$ with unit length so that they describe $2N(N-1)$ independent variables.
Then we use the fields $m_i$ to construct several special $SO(2N)$ covariant one-forms, that are orthogonal to $m_i$ and determine a basis of roots in $SO(2N)$.
Consequently, the combination of the fields $m_i$ and the covariant one-forms constructed from $m_i$ yields a complete basis states for $SO(2N)$ Lie algebra.
All together, they will be used to decompose the generic $SO(2N)$ connections. 
In the concluding part of the letter, we discuss the generalization of Abelian decomposition for $SO(2N+1)$ Yang-Mills theory.
It is straightforward provided the decomposition of $SO(2N)$ theory has been established. 

The $SO(2N)$ Lie group is rank $N$ and its Lie algebra has $N(2N-1)$ generators.
We denote them by $T_{a,b}$ with antisymmetric property, i.e.,
\( T_{a,b} = -T_{b,a} \) for \( a,b = 1~{\rm to}~2N \).
The generators are chosen in the defining representation 
as follows,  
\be
\left[ T_{a,b} \right]_{c,d} = -\frac{i}{\sqrt{2}} 
\left( 
\delta_{ac} \delta_{bd} - \delta_{bc} \delta_{ad} 
\right).
\la{Tab}
\ee
Here, we have normalized the generators such that 
\( {\rm Tr} (T_{a,b} T_{c,d}) 
= \delta_{ac}\delta_{bd} - \delta_{ad}\delta_{bc} \).
The commutation relations among the generators (\ref{Tab}) are easily obtained
\be
\left[ T_{a,b}, T_{c,d} \right] = \frac{i}{\sqrt{2}}
\left( \delta_{ac} T_{b,d} + \delta_{bd} T_{a,c}
      -\delta_{ad} T_{b,c} + \delta_{bc} T_{a,d}
\right).
\la{commutator}
\ee
This is the form for the rotational generators in a $2N$ dimensional real vector space.  

We designate the basis of the commuting generators in the Cartan subalgebra as
\be
H_{2i-1,2i} = T_{2i-1,2i},
\la{Hi}
\ee
where \( i = 1~{\rm to}~N \). 
Note that \( \left[ H_{2i-1,2i}, H_{2j-1,2j} \right]=0 \).
In terms of the generators $T_{a,b}$, a generic Lie-algebra element $v$ has the expansion \( v = \frac{1}{2} v^{a,b} T_{a,b} \). 
The factor one-half is needed to avoid double counting in the contraction summation.

Following \cite{fad2}, we now conjugate the elements of Cartan subalgebra $H_{2i-1,2i}$ (\ref{Hi}) by a generic element \( g \in SO(2N) \).
This gives $N$ Lie-algebra valued vector fields. 
They are 
\be
m_i = gH_{2i-1,2i}g^{-1} = \frac{1}{2} m_i^{a,b}T_{a,b}.
\la{mi}
\ee
Note that the fields $m_i$ remain invariant if $g$ transforms by a right diagonal factor \( g \to gh \), with $h$ belongs to the maximal Abelian subgroup of $SO(2N)$. 
In this way, $m_i$ produce an over-determined set of coordinates on the orbit \( SO(2N)/U(1)^N \) and depend on only $2N(N-1)$ independent variables.
In addition, they are orthonormal
\be
\left( m_i, m_j \right) \equiv {\rm Tr} \left( m_i m_j \right)
= \frac{1}{2} m_i^{a,b}m_j^{a,b} = \delta_{ij}.
\la{orthonormal}
\ee
Using (\ref{mi}), it is straightforward to verify that
\ba
\left[ m_i, m_j \right] & = & 0, 
\la{mimj} \\
{\rm Tr} \left( m_i \, dm_j \right) & = & \left(m_i, dm_j \right) = 0,
\la{midmj}
\ea
where \( dm_j = \partial_\mu m_j dx^\mu \).

Next, we proceed to consider an arbitrary Lie-algebra element $v$ under an infinitesimal adjoint action on the fields $m_i$.
We define this action by
\be
\delta^i v = \left[ v, m_i \right].
\la{adjointaction}
\ee
Applying the action $\delta^i$ twice and summing over the index $i$, we obtain a projection operator to a subspace which is orthogonal to the maximal torus and is spanned by the Lie-algebra valued fields $m_i$,
\be
(\delta^i)^2 v = v - m_i (m_i, v).
\la{delta2}
\ee
Note that the subspace in which (\ref{delta2}) projects corresponds to the space $SO(2N)/U(1)^N$, i.e., the roots of $SO(2N)$.
To derive (\ref{delta2}), we make use of (\ref{commutator}), (\ref{orthonormal}), and this equation
\[
\sum_i 
\left[ \left[ {\tilde v}, H_{2i-1,2i} \right], H_{2i-1,2i} \right]
= {\tilde v} - {\tilde v}^{2i-1,2i}H_{2i-1,2i},
\]
where \( {\tilde v} = g^{-1}vg \).

Having presented the basic formulas needed, we hope to generalize the method of Abelian decomposition for $SO(2N)$ Yang-Mills theory.
Introducing the matrix notation for the $SO(2N)$ connection one-form
\be
A = A_\mu dx^\mu 
  = \frac{1}{2} A_\mu^{a,b} T_{a,b} dx^\mu,
\la{Aym}
\ee
we parameterize this connection one-form $A$ into the following expression
\be
A = C^i m_i + \frac{1}{i} \left[ dm_i, m_i \right] + 
({\rm covariant~part}).
\la{Aoneform}
\ee 
The combination of the first two terms on the right-hand-side of (\ref{Aoneform}) is the so-called Cho connection, which was first introduced as a consistent truncation of the full four-dimensional connection \cite{cho2}.
It can be shown that, under $N$ independent gauge transformations generated by the Lie-algebra elements $\alpha^i m_i$, the Cho connection retains the full non-Abelian gauge degrees of freedom, while the one-forms $C^i$ transform as $U(1)$ connections, \( C^i \to C^i + d\alpha^i \).
Hence, the remaining part on the right-hand-side of (\ref{Aoneform}) (covariant part) must transform covariantly under gauge transformations and by construction must be orthogonal to the fields $m_i$. 
 
Because the decomposition method introduced by Faddeev and Niemi is on-shell complete, the number of field multiplets that appear in the decomposed connection (\ref{Aoneform}) have to be equal to that of physically relevant field degrees carried by the original $SO(2N)$ connection. 
It is known that the $SO(2N)$ Yang-Mills connection (\ref{Aym}) contains $2N(2N-1)$ physical components.
On the contrary, the Cho connection in (\ref{Aoneform}) introduces $N$ $U(1)$ connections $C^i$ and $N$ vector fields $m_i$.
The former contributes $2N$ physical degrees of freedom, while the later describes $2N(N-1)$ independent variables.
Adding both contributions up gives $2N^2$. 
As a result, the difference in degrees of freedom between both connections is
\be
2N(2N-1) - 2N^2 = 2N(N-1).
\la{degrees}
\ee  
This is the number of independent variables held exactly by the (covariant part) of (\ref{Aoneform}).
So, the space of (covariant part) is $2N(N-1)$ dimensional.
Moreover, according to the definition the fields appearing in the (covariant part) of (\ref{Aoneform}) are orthogonal to the fields $m_i$.
We thus deduce that the space of the (covariant part) coincides with the subspace to which the operator (\ref{delta2}) projects, the orbit $SO(2N)/U(1)^N$. 

In the following paragraphs, we shall use the fields $m_i$ to construct certain special Lie-algebra valued one-forms,  which determine the local basis of the (covariant part) space.  
What are these Lie-algebra valued one-forms? 
They can be gotten by repeatedly using the adjoint action (\ref{adjointaction}).
For instance, we first learn from (\ref{midmj}) that the Lie-algebra valued one-forms $dm_i$ are orthogonal to $m_k$. 
Let's denote $dm_i$ by $X_i$ for the purpose of later convenience and identify the one-forms $X_i$ as one subset of the basis states of the (covariant part) space. 
Next, we apply the adjoint action (\ref{adjointaction}) on $X_i$ to obtain another one-forms $Z_{ij}$,
\be
Z_{ij} \equiv \delta^j X_i = \left[ X_i, m_j  \right].
\la{dXi}
\ee
It is not difficult to see that, by utilizing (\ref{mimj}), the one-forms $Z_{ij}$ are orthogonal to $m_k$, too.
Hence, the one-forms $Z_{ij}$ can also be used to parameterize the basis of the (covariant part) space. 
Consequently, we continue to find the remaining one-forms that span the (covariant part) space by recurrent applying the adjoint action (\ref{adjointaction}) on the latest generated one-forms. 
After a little manipulation, we have
\ba
\delta^k Z_{ij} & = & \frac{1}{2} \delta_{ij}
                      \left[ \frac{1}{2}
                      \left( \delta_{ik} + \delta_{jk} \right) X_k + 
                      V_{ik} + V_{jk} 
                      \right] +
                      \delta_{ik}V_{kj} + \delta_{jk}V_{ki}, 
\la{dZij} \\
\delta^k V_{ij} & = & \frac{1}{2} 
                       \delta_{ik}Z_{kj} - \delta_{jk}U_{ki}, 
\la{dVij} \\
\delta^k U_{ij} & = & \frac{1}{4} \delta_{ij}
                      \left[ \frac{1}{2} 
                      \left( \delta_{ik} + \delta_{jk} \right) X_k +
                      V_{ik} + V_{jk}
                      \right] - 
                      \frac{1}{2}
                      \left(
                      \delta_{ik}V_{jk} + \delta_{jk}V_{ik}
                      \right).
\la{dUij}
\ea
It turns out that we get four subsets of Lie-algebra valued one-forms \( (X_i, Z_{ij}, V_{ij}, U_{ij}) \) in total, which form a closed algebra under the adjoint action (\ref{adjointaction}). 
The details of these one-forms are separately given in the Appendix.

The one-forms \( (X_i, Z_{ij}, V_{ij}, U_{ij}) \) possess definite properties under $SO(N)$ symmetries, for $N$ specifies the rank of $SO(2N)$.
See the Appendix for details.
For example, the one-forms $X_i$ yield the $SO(N)$ vector representation, $V_{ij}$ the $SO(N)$ rank-two tensor representation, and $Z_{ij}$ and $U_{ij}$ the $SO(N)$ symmetric tensor representations.
However, not all of the components in the one-forms $V_{ij}$ and $U_{ij}$ are independent.
It is shown, in the Appendix, that the rank-two tensor $V_{ij}$ satisfies two sets of constraints: \( \sum_i V_{ij} = \frac{1}{2} X_j \) and \( V_{ii} = 0 \) (no summation), and that the symmetric tensor $U_{ij}$ obeys also two sets of constraints: \( \sum_i U_{ij} = 0 \) and \( U_{ii} = \frac{1}{2} Z_{ii} \) (no summation).

After all, this enables us to count the number of independent components possessed by each one-forms $X_i$, $Z_{ij}$, $V_{ij}$, and $U_{ij}$.  
The dimension of the vector $X_i$ is $N$ and the dimension of the symmetric tensor $Z_{ij}$ is $\frac{1}{2}N(N+1)$.
Analogously, after taking the constraint equations into account, the dimension of the second rank tensor $V_{ij}$ is \( N^2-2N \) and the dimension of the other symmetric tensor $U_{ij}$ is \( \frac{1}{2}N(N+1)-2N \). 
The sum of these four numbers is $2N(N-1)$, which as expect coincides with the dimension of the space $SO(2N)/U(1)^N$.

As a result, \( (m_i, X_i, Z_{ij}, V_{ij}, U_{ij}) \) yields a complete set of basis states for the $SO(2N)$ Lie algebra, and can be used to decompose generic $SO(2N)$ connections.
To complete the decomposition, we need appropriate dual variables that appear as coefficients.
We observe that the Yang-Mills connection $A$ in (\ref{Aym}) is an $SO(2N)$ Lie-algebra valued one-form and transforms in the scalar representation of the $SO(N)$ group.
Accordingly, the variables that are dual to the one-forms \( (X_i, Z_{ij}, V_{ij}, U_{ij}) \) are undoubtedly zero-forms.
Let us denote them by \( (\phi^i, \psi^{ij}, \sigma^{ij}, \rho^{ij}) \), respectively.
These dual variables must transform in the same $SO(N)$ representations as the associated one-forms in order to form invariant combinations.\footnote{ 
The $U(1)$ connection one-forms $C^i$ in (\ref{Aoneform}) are the dual variables to the zero-forms $m_i$.
Thus, the $SO(N)$ group acts on the combination $C^i m_i$ trivially.}
  
We therefore conclude that the following decomposition of the four-dimensional $SO(2N)$ connection contains the correct number of independent variables, which are appropriate for describing the theory in the low-energy limit,  
\be
A = C^i m_i + \frac{1}{i} \left[ dm_i, m_i \right] + 
       \phi^i X_i + \psi^{ij} Z_{ij} +
       \sigma^{ij} V_{ij} + \rho^{ij} U_{ij}.
\la{Adecomposed}
\ee
According to the discussion of the Appendix, (\ref{Adecomposed}) can also be expressed in a gauge equivalent form
\be
{\tilde A} = 
\left( C^i - \frac{1}{i}R^{2i-1,2i} \right) H_{2i-1,2i} + 
\phi^i x_i + \psi^{ij} z_{ij} +
\sigma^{ij} v_{ij} + \rho^{ij} u_{ij},
\ee
where \( \left( x_i, z_{ij}, v_{ij}, u_{ij} \right) \) =
\( g^{-1} \left( X_i, Z_{ij}, V_{ij}, U_{ij} \right) g \).

The Wilsonian renormalization group argument suggests that, in terms of the field variables of the decomposed connection (\ref{Adecomposed}), the infrared $SO(2N)$ Yang-Mills theory takes the form
\be
S(m_i) = \int d^4 x
\left[ 
\left( \partial_\mu m_i \right)^2 +
\frac{1}{e_i^2} 
\left( \left[ \partial_\mu m_i, \partial_\nu m_i \right] \right)^2
\right].
\la{Saction}
\ee
The action (\ref{Saction}) is in the same universality class as that obtained in \cite{fad2} and is expected to describe stable, soliton-like configurations with nontrivial topological numbers \cite{fad1, fad2}.
It is interesting to investigate the detailed structures of the action.

In conclusion, we briefly summarize how the method of Abelian decomposition generalizes for $SO(2N+1)$ Yang-Mills connection. 
The $SO(2N+1)$ Lie algebra is rank $N$ and has $N(2N+1)$ generators.
As usual, we denote the generators by $T_{a,b}$ for $a,b$ = $1$ to $N+1$.
The $N$ Lie-algebra valued vector fields $m_i$ are constructed similar to (\ref{mi}) except that they depend on $2N^2$ independent variables.
The $SO(2N+1)$ connection one-form can still be parameterized by (\ref{Aoneform}), but the dimension of the space of the (covariant part) is $2N^2$.
Needless to say, the (covariant part) space is identical to the space $SO(2N+1)/U(1)^N$. 
The set of Lie-algebra one-forms, which determine the local basis of the space $SO(2N+1)/U(1)^N$, are found to be \( (X_i, Z_{ij}, V_{ij}, U_{ij}) \).
At this time, one-forms $V_{ij}$ fulfill one set of constraint instead of two.
It is \( V_{ii} = 0 \) (no summation).
As regards the one-form $U_{ij}$, the set of constraint satisfied by it is \( U_{ii} = \frac{1}{2} Z_{ii} \) (no summation).
It ends up that the number of independent variables carried by the set of one-forms \( (X_i, Z_{ij}, V_{ij}, U_{ij}) \) is $2N^2$, matching the dimension of the (covariant part) space.
Therefore, the expression (\ref{Adecomposed}) for the connection one-form is perfectly applicable to the decomposition of $SO(2N+1)$ Yang-Mills theory.  

The author is grateful to C. R. Lee for useful discussions.
This work was supported in part by Taiwan's National Science Council Grant No. 89-2112-M-194-003.

\vskip 2.0cm

\noindent{{\Large {\bf Appendix}}}

\vskip 0.5cm
 
We explicitly give the Lie-algebra valued one-forms $X_i$, $Z_{ij}$, $V_{ij}$, and $U_{ij}$, expanded in terms of the generators $T_{a,b}$ (\ref{Tab}). 
In (\ref{Adecomposed}), these one-forms are used to parameterize the local basis of the orbit $SO(2N)/U(1)^N$.

To begin with, we introduce the Maurer-Cartan one-forms
\be
L = dgg^{-1} \quad {\rm and} \quad R = g^{-1}dg,
\la{MConeform}
\ee  
then use (\ref{mi}) and (\ref{MConeform}) to rewrite  
\ba
dm_i 
& = & \left[ L, m_i \right] 
  = g \left[ R, H_{2i-1,2i} \right] g^{-1}, \\
\left[ dm_i, m_i \right] 
& = & g \left( R - R^{2i-1,2i}H_{2i-1,2i} \right) g^{-1}
  = L - g \left( R^{2i-1,2i}H_{2i-1,2i} \right) g^{-1}.
\ea
Because the (covariant part) of (\ref{Aoneform}) transforms covariantly under gauge transformation, we further represent the connection one-form (\ref{Aoneform}) in a form of manifestly gauge equivalent expression
\be
A = g \left[ 
\left( C^i - \frac{1}{i} R^{2i-1,2i} \right) H_{2i-1,2i} 
+ ({\rm c.p.}) \right] g^{-1}
+ \frac{1}{i} dgg^{-1},
\la{Aequivalent}
\ee
where \( ({\rm c.p.}) = g^{-1} ({\rm covariant~part}) g \).
Similar to what we have shown on the local basis of the (covariant part) space in (\ref{Aoneform}), the space of (c.p.) in (\ref{Aequivalent}) is likewise spanned by four Lie-algebra one-forms \( (x_i, z_{ij}, v_{ij}, u_{ij}) \).
They are related to the one-forms \( (X_i, Z_{ij}, V_{ij}, U_{ij}) \) of the (covariant part) as follows.

The $SO(N)$ vector one-form $x_i$ is defined by \( x_i = g^{-1} X_i g \), with
\be
x_i  = 
\frac{i}{\sqrt{2}} 
\left( R^{2i,a}T_{2i-1,a} - R^{2i-1,a}T_{2i,a} \right).
\la{xi}
\ee
Similarly, after introducing the set of identity \( \left( z_{ij}, v_{ij}, u_{ij} \right) \) = \( g^{-1} \left( Z_{ij}, V_{ij}, U_{ij} \right) g \), we have
\ba
z_{ij} 
& = & \frac{1}{2}
\Bigg[
\delta_{ij} 
\left( R^{2i,a}T_{2j,a} + R^{2i-1,a}T_{2j-1,a} \right) \non \\  
& + & R^{2i,2j-1}T_{2i-1,2j} + R^{2i-1,2j}T_{2i,2j-1} 
  - R^{2i,2j}T_{2i-1,2j-1} - R^{2i-1,2j-1}T_{2i,2j} 
\Bigg], 
\la{zij} \\
v_{ij} 
& = & \frac{i}{2\sqrt{2}} 
\bigg(
R^{2i,2j}T_{2i,2j-1} + R^{2i-1,2j}T_{2i-1,2j-1} \non \\
& - & R^{2i,2j-1}T_{2i,2j} - R^{2i-1,2j-1}T_{2i-1,2j} 
\bigg), 
\la{vij} \\
u_{ij}
& = & \frac{1}{4}
\Bigg[
\delta_{ij} 
\left( R^{2i,a}T_{2j,a} + R^{2i-1,a}T_{2j-1,a} \right) \non \\
& - & R^{2i,2j}T_{2i,2j} - R^{2i,2j-1}T_{2i,2j-1} 
  - R^{2i-1,2j}T_{2i-1,2j} - R^{2i-1,2j-1}T_{2i-1,2j-1} 
\Bigg]. 
\la{uij}
\ea

It is apparent from (\ref{vij}) and (\ref{uij}) that not all the components of $v_{ij}$ and $u_{ij}$ are independent.
$v_{ij}$ (\ref{vij}) satisfies
\ba
\sum_i v_{ij} & = & \frac{1}{2} x_j, \\
v_{ii} & = & 0 \quad ({\rm no~summation}).
\ea
In the same vein, in (\ref{uij}) we find two sets of constraints fulfilled by \( u_{ij} \),
\ba
\sum_i u_{ij} & = & 0, \\
u_{ii} & = & \frac{1}{2} z_{ii} \quad ({\rm no~summation}).
\ea

\end{document}